\documentclass{pasj00}

\begin{document}

\SetRunningHead{Matsuoka et al.}{Simplified Picture of LMXB}
\Received{2012/03/05}
\Accepted{2012/10/09}
\Published{2013/04/15}

\title{Simplified Picture of Low Mass X-ray Binaries based 
on Data from Aql~X-1 and 4U~1608$-$52}

\author{
Masaru \textsc{Matsuoka},\altaffilmark{1}
and
Kazumi \textsc{Asai} \altaffilmark{1} }
\altaffiltext{1}{MAXI team, RIKEN, 2-1 Hirosawa, Wako, Saitama 351-0198}
\email{matsuoka.masaru@riken.jp}

\maketitle

\begin{abstract}

  We propose a simplified picture of low mass X-ray binaries containing
a neutron star (NS-LMXBs) based on  data obtained from Aql~X-1 and
4U~1608$-$52 which often produce outbursts.
In this picture we propose at least three states and three state  
transitions; i.e., the states:  (1) soft state,  (2) hard-high state, and 
(3) hard-low state, and the state transitions:  (i) hard-high state to 
soft state, (ii) soft state to hard-high state, and (iii) hard-high  
state to hard-low state or vice versa.
Gases from the accretion disc of an NS-LMXB penetrate almost
the entire magnetic field and accrete onto the neutron star in cases 
(1) and (2), whereas in case (3) some gases accrete around the magnetic
poles in a manner resembling the behavior of an X-ray pulsar, and 
considerable gas is  
dispersed or ejected by the propeller effect. 
Transition  (iii) occurs when the Alfv\'{e}n radius is 
equal to the co-rotation radius. 
Therefore, in this case, it is possible to estimate the strength of
the neutron star's magnetic field  by detecting  transition 
(iii).  We also discuss the no-accretion X-ray state or  
recycled pulsar state, in which the Alfv\'{e}n radius  is larger than 
the light cylinder radius.
\end{abstract}

\section{Introduction}
A low mass X-ray binary (LMXB) is a binary star system comprising a neutron
star (NS-LMXB) or a black hole (BH-LMXB) with a late type 
companion star.  NS-LMXBs are known to emit X-rays  erratically  
and exhibit certain physical states by their emission behavior, 
which is complicated owing to the different  natures of the sources.
Therefore, no unified model or picture of NS-LMXBs has been established since
the discovery of Sco~X-1, considered a prototype of NS-LMXBs, in 1962
(Giacconi et al. 1962).
An NS-LMXB is generally accompanied by an old neutron star with a weak magnetic
field. Thus, the behavior of an NS-LMXB is similar to that of a BH-LMXB
in variability and state transitions; however,
the existence of a rigid surface and an intrinsic magnetic field as well as
smaller gravitational forces in NS-LMXBs differentiates them from BH-LMXBs.
With this in mind we propose a simplified picture of NS-LMXBs based on the data
from Aql~X-1 and 4U~1608$-$52, which often produce outbursts
(e.g., Asai et al. 2012 and references therein).
Further, our goal is to provide a picture that is also consistent with
results concerning NS-LMXBs in general.

Since the discovery of Sco~X-1 in 1962 (Giacconi et al. 1962),
approximately 100  NS-LMXB sources have been discovered
(see, e.g. the catalogue by Liu, van Paradijs \& van den Heuvel 2007) using X-ray 
astronomy satellites. However, we do not have a complete picture of NS-LMXBs.
Depending on their  luminosity, their spectra  include 
blackbody radiation, non-thermal Comptonization, and other components that
obey a power law function with a cut-off energy.  Weak thermal or 
fluorescent iron emission  lines are also detected.
The variability of the NS-LMXB intensities reveals both quasi-periodic
oscillations (QPOs) of various periods in addition to erratic variation
and coherent milli-second pulsations for some NS-LMXBs
(e.g., Liu et al. 2007; Archibald et al. 2009).
However, the pulsation behavior differs from that of ordinary X-ray binary 
pulsars which are generally  binary systems consisting of a younger  
neutron star and an early type star.

Twenty-two years after the discovery of Sco~X-1, the Japanese X-ray astronomy
satellite Tenma provided a breakthrough in the understanding of the X-ray
spectral structure of NS-LMXBs  (Mitsuda et al. 1984, 1989).
The satellite observation showed that the X-ray spectra of Sco~X-1,  
GX~5$-$1, and other bright NS-LMXBs are reproduced well by a blackbody 
spectrum of $kT\sim$2~keV and a multi-color disc model spectrum with
$kT_{\rm in}\sim$1~keV (inner disc temperature).

Since a simplified Mitsuda model (Mitsuda et al. 1984; a prototype of 
the so-called Eastern model)
could not be applied to hard-state NS-LMXBs, the so-called Western model for
LMXBs was proposed (White, Stella \& Parma 1988).
The prototype of the Eastern model was also modified by factoring in
the effects of Comptonization (Mitsuda et al. 1989).
Both the Western and Eastern models account for the hot plasma
that can create hard spectra containing a power law
spectrum, a broken power law spectrum, or a Comptonized spectrum.
The current X-ray spectrum for the hard state of NS-LMXBs is explained by
introducing the model parameters of seed photons and high energy electrons
for Comptonization, but the physical meaning of these parameters is not completedly
justified in these models.

In this study, we account for the differences in physical 
structure that lead to the difference between  the X-ray spectra of
NS-LMXBs for the hard and soft states. 
Moreover, we note that the broad band spectra observed with Beppo-SAX,
RXTE and Suzaku are described by a model consisting of a thermal soft component and
a Comptonized hard component that extends to several tens of keV
(Barret 2001; Gierli\'{n}ski \& Done 2002; Lin, Remillard \& Homan 2007; 
Sakurai et al. 2012). Recently, a new model to account for the effects of  
both thermal and dynamical Comptonization has been proposed; it  fits  
the data from the different spectral states obtained by the BeppoSAX,  
INTEGRAL and RXTE satellites (Farinelli et al. 2008).
However, in this paper, we do not aim to consider detailed spectral models:
our goal is primarily to obtain a simplified physical picture in 
the region near the neutron star and describe the 
underlying configurations that would pertain to the different 
states of NS-LMXBs.

 Thus far, the spectral features of NS-LMXBs have been classified using
color--color diagrams and hardness--intensity diagrams
(e.g., Hasinger \& van der Klis 1989, Schulz, Hasinger \& Truemper 1989).
These diagrams are the results of phenomenological investigations of
NS-LMXBs (i.e., event-based studies);  however, factors such as the  
geometrical structure and emission mechanisms have not been clearly explained
using  these diagrams.
Furthermore, some NS-LMXBs display quasi-periodic oscillations (QPOs), 
with various modes at certain times (Hasinger \& van der Klis 1989).
Approaches based on QPOs have also been used to classify LMXBs; 
however, they  have not led to a clear underlying
physical picture. Therefore, we do not follow the paradigm 
of classifying NS-LMXBs using color-color and  hardness-intensity 
diagrams, but simply consider the different states 
and state transitions according to the X-ray intensity 
or  mass accretion rate. 

In this paper, we investigate all the physical features of NS-LMXBs, 
including their X-ray intensity and spectral evolution from high  
luminosity in the soft state to extremely low luminosity in the hard state.
In this scenario, the magnetic fields of NS-LMXBs play
an important role, and the magnetic pressure is comparable to
the accretion gas pressure in local regions as well as on the whole.
These pressures uniquely affect the behavior of NS-LMXBs, because they  
play significant roles in each state transition.
Furthermore, although significantly different models often provide 
acceptable fits to the same data (Lin, Remillard \& Homan 2007),  
we consider two major components to be necessary;  one or more 
blackbody components and the Comptonized component, which is 
produced by high energy electron gases and seed photons from the
 blackbody components. Other components such as iron line 
emissions or a power law component play  special roles in some
 cases.

Our simplified picture is based on long-term MAXI data for Aql~X-1 and  
4U~1608$-$52, whose luminosities are less than the $\sim 0.3$ Eddington limit
(Asai et al. 2012); however, because the MAXI data (Matsuoka et al. 2009 for MAXI) 
were found to be inadequate, it was supplemented by satellite data from  
RXTE, Asca, Beppo-SAX, Swift, INTEGRAL, Suzaku, Chandra, XMM-Newton, and others.  
We also employ the recycled NS-LMXB  pulsar scenario, 
although this has not been  completely confirmed (e.g.,Archibald et al. 2009; Ho, 
Maccarone \& Andersson 2011).  In section~2, the three states and three  
state transitions involving NS-LMXBs
are summarized, and we describe the physical processes and 
features of each state and each transition. Furthermore, we 
investigate the case of  no-accretion case,  which
is considered to be strongly correlated with a recycled pulsar.
In subsection~3.1, we consider the different X-ray spectra, which  
depend on the viewing angle.  In subsection 3.2, we attempt to examine the correlation
between our picture and  classification using color--color and 
hardness--intensity diagrams of NS-LMXBs.
Finally, in subsection~3.3, we summarize our picture.

\section{Physical processes and features of NS-LMXB states and state transitions}

\subsection{Physical parameters for states and transitions}

We begin by describing  some physical parameters that are employed 
in this investigation.
First, the co-rotation radius $R_{\rm c}$ for a neutron star  
having  spin period $P$  is expressed as    
\begin{eqnarray}
R_{\rm c} &=& (GMP^2 / 4\pi^2)^{1/3} \nonumber \\
&=& 1.7 \times 10^6 (M/1.4 \Mo)^{1/3} (P/1~\rm{ms})^{2/3}~ {\rm cm},
\end{eqnarray}
where $G$, $M$ and \Mo ~ are the gravitational constant, the mass of the
neutron star, and  solar mass, respectively. 
Next, we derive the Alfv\'{e}n radius, which depends on the mass accretion rate
as well as the magnetic field, mass and radius of the neutron star.
Assuming that the mass flow through the accretion disc from a companion
star is spherically symmetric, the Alfv\'{e}n radius ($R_{\rm A0}$) in which the gas pressure
is equal to the magnetic pressure on a magnetosphere is expressed as
\begin{eqnarray}
R_{\rm A} = \eta R_{\rm A0} = 3.7 \times 10^6  \eta (L/10^{36} {\rm erg~s^{-1}})^{-2/7}(B/10^8 {\rm~G})^{4/7} \nonumber \\ 
\times (M/1.4\Mo)^{1/7}(R_{\rm ns}/10^6 {\rm cm})^{10/7}~ {\rm cm},
\end{eqnarray}
where the original expression (27)  of Ghosh and Lamb (1979a) is changed by substituting
$\mu=BR_{\rm ns}^3$ and  $dm/dt = LR_{\rm ns}/GM$ for the equation of the 
magnetic dipole moment of the neutron star and the relation
between the mass accretion rate $dm/dt$ and
the total luminosity $L$, respectively.
Here, $B$ and $R_{\rm ns} $ are the magnetic field and  radius of
the neutron star, respectively.
Note that the energy release due to $dm/dt$ is
transfered to radio jets or  plasma jets as well as X-rays.  The factor $\eta $ is 
expressed as a value of 1 $\sim $ 0.52 depending on the model and the  effect of
the mass accretion flow   (Ghosh  \&  Lamb 1977; 
Elsner  \&  Lamb 1977; Ghosh  \& Lamb 1979a,b).

The accretion gases under the condition of $R_{\rm A} < R_{\rm c}$
penetrate to the neutron star's surface through the intrinsic magnetic field.
Thus, if the transition point of  $R_{\rm A}=  R_{\rm c}$  is detected, 
we can obtain the magnetic field of the neutron star
\begin{eqnarray}
 B = 2.6 \times 10^7 \eta^{-7/4}  (P/1~{\rm ms})^{7/6} (L/10^{36} {\rm erg~s^{-1}})^{1/2} \nonumber \\
\times (M/1.4\Mo)^{1/3}(R_{\rm ns}/10^6~{\rm cm})^{-5/2}~{\rm G},      
\end{eqnarray}
where the model dependence factor, $\eta^{-7/4}$,  is expressed in a value of 1 $\sim$ 3.
Here, we consider the physical behavior with respect to  $R_{\rm A}$ 
and  $R_{\rm c}$.  The condition of $R_{\rm A} < R_{\rm c}$  or
$R_{\rm A} \ll R_{\rm c}$
is realized around the inner local region of the thin disc or the 
region of gas falling from the disc as well as under a strong gas flow.
If the gas transferred from the companion star is concentrated
from a thick disc to a thin disc, the gas pressure increases
by $\sim H_{\rm thick} / H_{\rm thin} > 1$
in the local boundary region where $H_{\rm thick}$ and $H_{\rm thin}$
are the thicknesses of the thick and thin discs, respectively.
This is a reasonable situation, because the amount of accreted
gas does not change very rapidly just before and after the transition.
On the other hand the condition of $R_{\rm A} > R_{\rm c}$  or
$R_{\rm A} \gg R_{\rm c}$   will cause a propeller effect for
accreting gas.  This means that the gases reaching the magnetosphere
will be rejected by this effect, but some part of the gas will accrete
onto the polar region if the accreting gases are distributed widely.

Finally, the Alfv\'{e}n radius becomes comparable to the radius of
the light cylinder whose value is given by
$R_{\rm LC} = Pc/2\pi$,
where $c$ is the velocity of light.
When this occurs, gases from the accretion disc cease accretion onto the neutron star.
Thus, we call this a recycled pulsar state in which the accretion 
X-rays will  not be produced.
If  this turning point is detected, the neutron star's  magnetic field can be 
estimated from the following relation when  $R_{\rm LC} = R_{\rm A}$:
\begin{eqnarray}
 B = 1.6 \times 10^7 \eta^{-7/4} (P/1~{\rm ms})^{7/4} (L/10^{34} {\rm erg~s^{-1}})^{1/2} \nonumber \\
\times (M/1.4\Mo)^{-1/4}(R_{\rm ns}/10^6~{\rm cm})^{-5/2}~{\rm  G},      
\end{eqnarray}
where  a dipole magnetic field and a uniform accretion rate are assumed.
The magnetic field thus obtained for a certain source would then be the same as that 
given by 
equation~(3) if it does not decay very much. 
Note that it is genarally difficult to  determine $L$, because of 
the need to adopt  the total mass accretion rate, i.e., because
we have to add the considerable luminosity of the ejected plasma
to the X-ray luminosity.  The gases to be accreted below this turning 
point will be dispersed by the propeller effect in the magnetosphere  
of the spinning neutron star.

\subsection{Summary of the present simplified picture of NS-LMXBs: 
Classifications of  states and transitions}

\begin{figure*}
  \begin{center}
    \FigureFile(130mm,){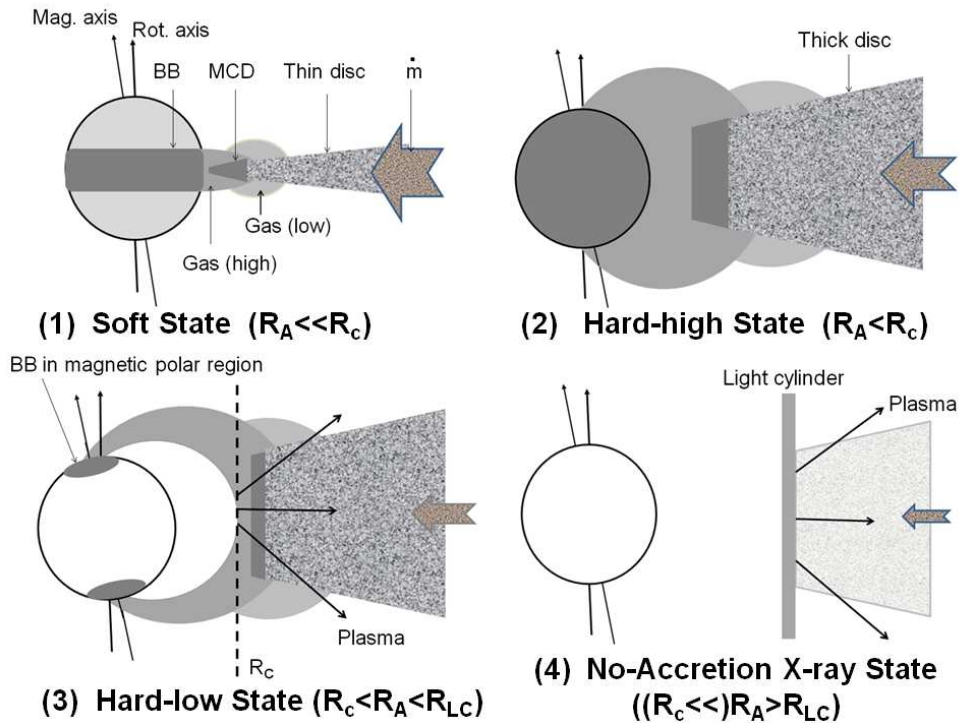}
  \end{center}
  \caption{
Conceptual cartoons for four states of an NS-LMXB.
BB = black-body, Gas(high/low): gases with high/low energy electrons
responsible for Comptonization
(also see figure~4 and the text in subsection~3.1),
MCD = multi-colour disc, $R_{\rm A}$ = Alfv\'{e}n radius,
$R_{\rm c}$ = co-rotation radius,
$R_{\rm LC}$ = radius of light cylinder,
Thin/thick disc = geometrically thin/thick disc. 
\.{m}=$dm/dt$: mass accretion rate from a companion star.
In the white region, i.e., in  (3) and (4) on the neutron star, gas 
accretion is impossible, whereas in the gray region in (1), it  
is possible, but little gas is accreted. 
   }
\label{fig1}
\end{figure*}

In this section we investigate the physical processes and features 
of the three states and three transitions of NS-LMXBs.  The
explanation is supplemented by the conceptual cartoons
 shown in figure 1.  In the last subsection, 2.6, we  discuss
 the no-accretion X-ray state, or recycled pulsar state, in which 
weak  X-rays are likely produced even though the Alfv\'{e}n
 radius is larger than the light cylinder radius. 

\subsubsection{ Classification of physical states}

(1) {\bf Soft state:} Soft spectrum with high intensity.
  Physical state: $R_{\rm A} \ll R_{\rm c}$ in the inner region
of the accretion disc.
 A thin (i.e., geometrical thin with dense gas) accretion disc is formed. 

(2) {\bf Hard-high state:} Hard spectrum with high intensity.
  Physical state: $R_{\rm A} < R_{\rm c}$. A thick (geometrical thick with
density thin gas) accretion disc is formed.

(3) {\bf Hard-low sate:} Hard spectrum with low intensity.
  Physical state: $R_{\rm LC} > R_{\rm A} > R_{\rm c}$.  A thick accretion disc is formed.
Considerable amounts of accretion gas are ejected by the propeller effect, 
depending on the accretion rate.

(4) {\bf No-accretion X-ray state:} Physical state: 
 $R_{\rm A} > R_{\rm LC}$. 
Most accretion gases are rejected by the propeller effect.    
The NS-LMXB in this state may be going to a recycled pulsar.

\subsubsection{Classification of transitions between states}

(i) {\bf Transition from hard-high state to soft state:} $R_{\rm A} < R_{\rm c}$ 
to  $R_{\rm A} \ll R_{\rm c}$ in the inner region 
of the thin disc. This accompanies 
the disc transition from  thick to thin.  The transition luminosity 
is not defined absolutely, but depends on the pre-outburst 
luminosity (Asai et al. 2012).  

(ii) {\bf Transition from soft state to hard-high state:} $R_{\rm A} \ll R_{\rm c}$ 
to $R_{\rm A} < R_{\rm c}$. This accompanies
the disc transition from thin to thick.  The transition luminosity is 
1\%--4\%  of the Eddington luminosity (Maccarone 2003).
 
(iii){\bf Transition from hard-high to hard-low state or vice versa:} 
Transition luminosity is defined at the level of $R_{\rm A} = R_{\rm c}$. 

(iv) {\bf Transition from hard-low to no-accretion X-ray state:} 
This transition has not been confirmed in Aql X-1, 4U1608-52, 
or any other NS-LMXBs.  We define this transition point 
at $R_{\rm A} = R_{\rm LC}$, and hypothesize its occurrence towards 
recycled pulsars.

\subsection{Soft state}

\subsubsection{The transition from the hard-high to the soft state}

An increase in the mass flow rate from the companion star leads to a state
transition from the hard-high state to the soft state in the standard  
accretion disc that forms around an old neutron star. 
A state transition or instability in the accretion disc was initially 
indicated in  black hole binaries and dwarf novae
(e.g., Mineshige \& Wheeler 1989; Abramowicz et al. 1996).
The transition from a geometrically thick accretion disc to a thin
disc occurs because of a change in the disc, which depends on  the 
mass flow rate; thus, thermal radiation in the thin disc becomes 
dominant. Furthermore, most of the gases in the disc are accreted in a 
focused manner, towards the neutron star's  equatorial region.	
When this occurs, the gas pressure in the neutron star is much larger
than the magnetic pressure; 
that is, $R_{\rm A} \ll R_{\rm c}$.  This implies that the
gas pressure in the soft state is usually larger than both that
in the hard-high state and the magnetic pressure in the neutron star
(e.g., Campana et al. 1998b).

When an outburst occurs in an NS-LMXB, the rise time into the disc
transition from thick to thin (defined as the duration 
from the onset of the outburst to the hard-high to soft transition)
 depends on the luminosity of the NS-LMXB before the outburst. 
A large luminosity corresponds to a slow rise time, whereas a weak 
luminosity corresponds to a fast rise time.  Both fast and slow 
rise times are distinctly observed in the outbursts of Aql~X-1 and 
4U~1608$-$52 (Asai et al. 2012; Yu \& Yan 2009), although 
NS-LMXBs may occasionally exhibit a median type of rise time.
The two distinct rise times are explained by the fact that 
the disc transition is delayed
by energy input to the accretion disc due to irradiation (Asai et al. 2012).
This irradiation effect is similar to that in BH-LMXBs
(Gierli\'{n}ski \& Newton 2006). 
The effect of accretion disc irradiation in BH-LMXBs has
been investigated theoretically (e.g., Kim et al. 1999).
The delayed hard-to-soft transition 
 has been attributed to the hysteresis of
both BH-LMXBs and NS-LMXBs (Miyamoto et al. 1995; 
Maccarone \& Coppi 2003; Yu et al. 2004).
To further understand this behavior, it is necessary to conduct
magneto-hydro dynamical studies of disc instability in  NS-LMXBs,
as done by, e.g.,   Oda et al. (2010), for  BH-LMXBs.

\subsubsection{X-ray emission mechanism in the soft state}
In order to explain X-ray emission from bright NS-LMXBs Mitsuda 
et al. (1984) introduced a simple blackbody and multi-color 
blackbodies in the accretion disc. 
The simple blackbody 
is considered to be created in the equatorial region of the neutron 
star, with an accretion rate proportional to the belt width.  
It has also been defined recently as the boundary layer or the 
equatorial belt for the blackbody emission region in the soft state 
of 4U~1704$-$44 (Lin, Remillard \& Homan 2010). The equatorial belt 
model might generally be explained by the spectrum in the soft 
state of bright NS-LMXBs near the Eddington limit (i.e., 
Z-sources).  
   
The equatorial belt model is also applicable, in principle, to the soft 
state for Aql~X-1 and 4U~1608$-$52 (i.e., Atoll sources) , which have 
lower  Eddington luminosities (Mitsuda et al. 1989; Takahashi, Sakurai 
\& Makishima 2011; 
Sakurai et al. 2012). It is  natural  that the accretion gases 
are focused at the equatorial belt when 
$R_{\rm A} \ll R_{\rm c}$ with increasing of gas pressure due to
$ H_{thick} / H_{thin} > 1 $ as indicated  in the subsection 2.1.
However, the temperature and the width of the belt do not behave simply, 
as in the soft state of bright NS-LMXBs (i.e., Z-sources). 
The behavior in the soft state of 4U~1608$-$52 is investigated  
in detail by Takahashi, Sakurai \& Makishima (2011).  

On the other hand the radiation from the multi-color
blackbodies is also generally considered to be another major 
component in the soft state for Atoll sources as well as Z-sources.
The dense geometrically thin disc
radiates a multi-color spectrum; in particular, a soft spectrum with
$kT_{\rm in} \sim 1$~keV is radiated from its inner region 
 (Mitsuda et al. 1984, 1989).
The temperature of 
the multi-color inner disc depends somewhat on the sources and
the accretion rate (Takahashi, Sakurai \& Makishima 2011).  
Consequently, the actual fitted temperature values yiel  
a considerable range of values. 

The iron line emission and  Comptonized spectrum are added 
as minor contribution to the above two thermal spectra in the soft 
state.  As shown in figure 1 (1), the gases surrounding the neutron 
star may be able to produce these components. Furthermore, we can  
also expect the radiations from these gases by themselves, but 
appreciable amounts are not observed.  Therefore, when the 
accretion of a gas increases, radiation accompanying the 
outflowing gas will appear (Takahashi, Sakurai \& Makishima 
2011), which is not a minor component in the soft state beyond  
the Eddington limit (Homan et al. 2010).  Investigation 
of these components is, however, beyond the scope of this paper.

\subsubsection{Radio emission during hard-high to soft transition}

NS-LMXBs whose luminosities do not reach the Eddington 
limit such as Aql~X-1 and 4U~1608--52 probably do not produce observable  
radio jets in the soft state (Migliari \& Fender 2006 and references therein). 
This is a different behavior  from that of higher luminosity NS-LMXBs such as 
Sco~X-1  (Penninx et al. 1988; Hjehlming et al. 1990; Fender et al. 2007). 
On the other hand, a radio jet is occasionally observed in the hard-high 
state; two different cases are observed.  
As the first case, Miller-Jones et al. (2010) made simultaneous 
X-ray and radio observations of the outburst by Aql~X-1 in November 
2009.  A radio burst was detected for $\sim 5$~ days in the initial 
phase of the outburst, but this signal ceased after a specific 
time interval. On the other hand, X-rays were observed for $\sim 25$~days
(the entire period of the outburst); moreover, the transition from
the hard-high to the soft state occurred during this outburst.
The second case is observed in some NS-LMXBs that produce radio jets
without a clear state transition (Fender et al. 2004; Migliari \& 
Fender 2006).

It is speculated that the hard-high state has the potential of producing 
radio jets towards the soft transition with some instability in the 
pressure balance.  In the first case above, the condition of 
$R_{\rm A} \ll R_{\rm c}$ occurred owing to the state transition of
the disc; i.e., the gas pressure remained higher for a certain time,  
although a radio jet was produced.  In the second case, the 
condition of $R_{\rm A} \ll R_{\rm c}$ with a disc transition did not occur  
owing to the radio  outburst; i.e., the gas pressure was released because of  
the radio jet.  It is too complicated to predict which case occurs in an 
outburst, because this depends on combined factors of the mass  
flow rate from the  companion star and  the disc instability.

\subsection{Hard-high state}
Unlike the transition from the hard-high state to the soft state,
the transition from the soft state to the hard-high state is considered
to occur at 1\%--4\% of the Eddington luminosity, and this range seems 
to be common to both  NS-LMXBs and  BH-LMXBs (Maccarone 2003).
The luminosity value for the soft to hard state transition is lower
than that for the hard-to-soft state transition for a
slow rise time, as found by 
Asai et al (2012). As we already suggested in the previous section, 
this transition can be
caused  by disc instability due to the decrease in the mass flow rate
from the companion star (e.g., Mineshige \& Wheeller 1989; Abramowicz et al. 1995).

As mentioned above, the gas pressure in the soft state is substantially greater
than the magnetic pressure of the neutron star.
Therefore, gases will be accreted on the equator through the magnetic
field if the accretion disc is geometrically thin and optically  
thick. If an accretion
disc transition occurs from the thin disc of the soft state to the thick disc
of the hard state, the gas pressure in the geometrically thick disc decreases.
However,  gases from the thick disc accrete onto the entire surface of
the neutron star if the Alfv\'{e}n radius is less than the co-rotation
radius.  This accreted gas will create a blackbody that expands  
over the entire neutron star. 
The blackbody region with $kT\simeq0.3\sim$1~keV constitutes the seed
photons required for the Comptonization process.
The parameters for this blackbody region and the Comptonization can
be obtained by spectral fitting in the hard-high state
(e.g., Lin, Remillard, \& Homan 2010; Sakurai et al. 2012).  

In the hard-high state, the hot gases responsible for the Comptonization process
also cover almost the entire neutron star as accreted gases from the thick
accretion disc as shown in figure 1(2).
In this configuration, the NS-LMXB can often emit strong Comptonized spectrum
extending from several tens of keV to $\sim200$~keV
(Gierli\'{n}ski \& Done 2002; Lin, Remillard \& Homan 2007; Sakurai et al. 2012). 
This is consistent with the model fitting parameters for the hard-high state
which provide us an equivalent blackbody of $kT\simeq0.3\sim$1~keV
for a neutron star size for seed photons.
The X-ray emission for the entire neutron star also shows variations
in the QPO; this is because of the interaction between the accreting gases
and the intrinsic magnetic field in the rotating surface of the neutron star.
Thus, the X-ray spectrum and variability in the hard-high state 
differ greatly from those in the soft state.  

The existence and duration of the hard-high state both depend on the gas
accretion rate and the intrinsic magnetic field,
as given by the relation $R_{\rm A} < R_{\rm c}$.
In fact, the extended limit of the luminosity for a soft transition depends
on the disc transition which is influenced by the mass flow rate and
irradiation rate to the disc, whereas the transition luminosity for
the hard-high to hard-low state transition depends on both the mass  
accretion rate and the magnetic field.

\subsection{Hard-low state}
\subsubsection{Transition point between hard-high and hard-low states}

\begin{figure*}
  \begin{center}
    \FigureFile(150mm,){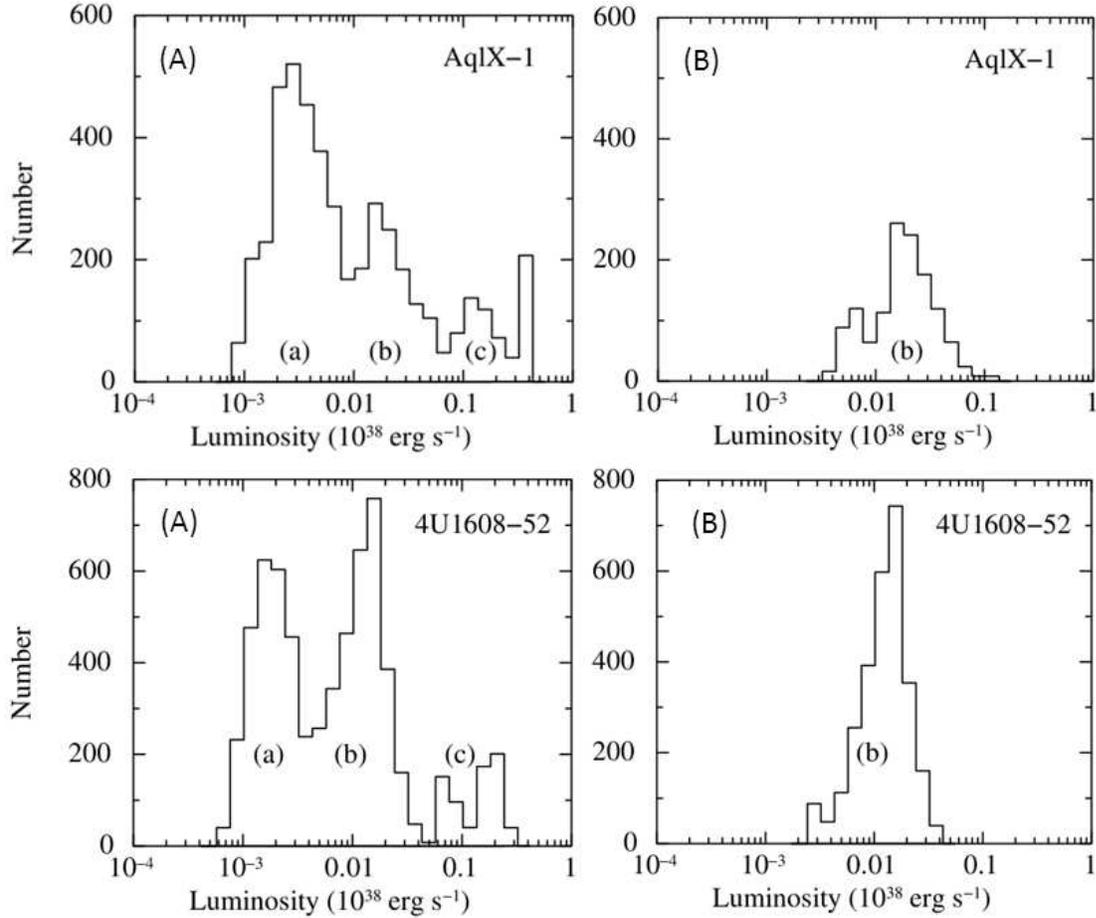}
  \end{center}
  \caption{Intensity distributions of Aql~X-1 and 4U~1608$-$52.  
 (A): All one day data above 1$\sigma $ from MAXI/GSC observed from 
August 15, 2009 to May 31, 2012 were adopted. 
(B): All one day data above 4$\sigma $ without the data  in the 
soft state in the same observational period as in (A). 
The flux per day is converted to the luminosity 
(erg~s$^{^-1}$) in the 2--10 keV band using the same method as Asai et al. 
(2012).  Three clear groups (a, b, and c) with peaks and  valleys are 
seen in (A) for each source.  See the text for the 
correspondence  with the respective states. 
}
\label{fig2}
\end{figure*}

A way to discriminate between the hard-high and hard-low states is proposed here for the 
first time. This detailed observational study is based on  MAXI/GSC and RXTE/ASM data 
for Aql~X-1 and 4U~1608$-$52, which will be published elsewhere (Asai et al. 
in preparation).  The transition point of these two states occurs at a luminosity 
corresponding to $R_{\rm A} = R_{\rm c}$.  Note that the X-ray luminosity 
around this point does not exactly coincide with the mass flow rate, because 
considerable gas will be ejected by the propeller effect as shown in figure 1(3). 
Therefore,  neglecting this effect would be an oversimplification.  
Here, we discuss this problem in terms of the following consideration.

In order to examine whether we can divide the hard state 
into two states, we create an intensity distribution from long-term 
data observed by MAXI/GSC as shown in figure 2.  The data points for each 
day are converted to a luminosity in the 2--10 keV band, which is obtained 
by the same method as that of  Asai et al. (2012).  
Each panel (A) on the left-hand side of figure 2 shows 
the distribution for Aql~X-1 or 4U~1608$-$52, where the public data were 
collected natively from August 15, 2009 to May 31, 2012. First, according to
the previous analysis (Asai et al. 2012), we understand that 
each (c) group corresponds to the soft state for two sources that are 
distributed above a luminosity of $(5 \sim 8) \times$ $10^{36}$ erg s$^{^-1}$.
Each valley between (b) and (c) also corresponds to a transition range 
from the hard-high to the soft state and vice versa. The structure of the 
distribution (c) indicates the spectral behavior in the soft state, 
where its detailed behavior is shown on two-dimensional diagrams  
for each outburst (Asai et al. 2012).

As shown in Asai et al (2012), the difference in luminosity level between 
the soft and hard state varies from burst to burst; however, it is possible to
discriminate states if we use the intensity ratio of Swift/BAT  
to MAXI/GSC (Asai et al. 2012) or RXTE/ASM (Yu \& Yan 2009). In this  
investigation, we are interested in the hard state 
and not the soft state.  Furthermore, we know that distribution (a) 
in each panel (A) in figure 2 is contaminated by the background. 
Therefore, we created a panel (B) for each of the two sources, in which the 
data points in the soft state were removed, furthermore, the points above the
4$\sigma $ level were employed for each source. 

Now we investigate the distributions on panel (B) in the right-hand side 
of figure 2. The 4$ \sigma $ 
levels correspond to $ \sim 4.0 \times 10^{35}$ erg s$^{^-1}$ for Aql~X-1 
and $ \sim 2.5 \times 10^{35}$ erg s$^{^-1}$ for 4U~1608--52.
Consequently, it is considered that the intensity distribution above 
these levels on the panel (B) is mostly provided with each source  
and that corresponds to the data in the hard state.  The last hard state 
luminosity in the hard to soft state transition and the first hard state 
luminosity in the soft to hard state transition are in the higher luminosity 
side of distribution (b).  The lowest luminosity of these transitions 
is around the 1 \% Eddington luminosity as determined by Asai et al. (2012) 
and also as indicated in  sub-subsection  2.5.3.

We consider the lower side of distribution (b).  
If gas accretion onto the neutron star continues at a certain 
level, distribution (b) should be extended to lower luminosity 
randomly. However, we can observe a cut-off (e.g.,``clear" 
cut-off for 4U~1608--52) in the lower luminosity 
side of each  distribution (b).  Two reasons for this cut-off are 
considered: (1) the propeller effect under   $R_{\rm A} > R_{\rm c}$, or
(2) stopping of the accretion gas in a certain region by some 
instability in the accretion disc.  
Although reason (2) may be a future issue for theoretical work, 
it is neccesary to keep gases at a certain region above 
the neutron star's surface if any accretion gas is not rejected. 
Thus, we prefer reason (1) to explain this lower luminosity cut-off. 
We discuss the behavior of the propeller effect in the following 
sub-subsections. 

If  $R_{\rm A} > R_{\rm c}$ is realized effectively, 
it is possible to discriminate between the two  hard states 
at the 
point of $R_{\rm A} = R_{\rm c}$: luminosities higher than 
this point correspond to the hard-high state, whereas lower 
luminosities correspond to the hard-low state.  
If the propeller effect occurs when 
$R_{\rm A} > R_{\rm c}$, the data population will decrease considerably. 
Although MAXI/GSC cannot detect the intensities lower than 
$ (2 \sim 3) \times 10^{35}$ erg s$^{^-1}$  for the two sources,  
X-rays with the luminosities of $ 10^{32 \sim 34}$ erg s$^{^-1}$ 
are often observed by the ASCA, Chandra, Suzaku and XMM-Newton satellites 
(e.g. Asai et al. 1996; Cackett et al. 2011).  

As we discuss in sub-subsection 2.5.5  and subsection 
2.6, it is not definitely confirmed whether the X-ray intensity in the
quiescent period ($  10^{32 \sim 34}$ erg s$^{^-1}$) belongs to the 
hard-low state or the no-accretion X-ray state.  However, the  
main objective here was to define the level of  $R_{\rm A}=R_{\rm c}$. 
Probably we have achieved this objective; thus, we can obtain the cut-off
luminosity  $ \sim 1.6 \times 10^{36}$ erg s$^{^-1}$ for Aql~X-1 
and 4U~1608--52 from each panel (B) in figure 2, where we tentatively
employed  the luminosity from the starting point in the lower 
luminosity cut-off of  distribution (b). Here, both coincidentally 
have the same value within uncertainty. 
A detailed quantitative investigation will be 
published elsewhere (Asai et al. in preparation).

\subsubsection{Transition to the hard-low state (X-ray pulsars)}

  When the Alfv\'{e}n radius $R_{\rm A}$ becomes larger than the co-rotation radius
$R_{\rm c}$ because of a decrease in the mass accretion rate, the gases near  
the neutron star's surface cannot penetrate entirely through the magnetic 
 field. When that occurs, considerable gas is dispersed or 
ejected by the propeller effect, although some of the gas does accrete around 
the magnetic poles, in a manner similar to that in an X-ray pulsar.
This picture is typical for understanding the evolutionary link
between NS-LMXBs with old neutron stars and X-ray pulsars with younger neutron
stars.  Thus far, a coherent pulsation (whch has been attributed to the 
spin of the neutron star) has been detected in several NS-LMXBs
(e.g., Liu, van Paradijs \& van den Heuvel 2007).  These pulsations are often 
observed at the time of X-ray bursts, indicating that the polar region may not
be solely responsible for them (see sub-subsection~2.5.4).

If the point of transition from the hard-high state to the hard-low state (or
vice versa) can be detected, it is possible to estimate the strength of
the neutron star's magnetic field  by using the accretion rate at
$R_{\rm A} = R_{\rm c}$.  Since the accreting gas is thought to cover the 
entire magnetosphere of the neutron star in both the hard-high and 
hard-low states of the NS-LMXB,
we can estimate the magnetic field from  equation~(3) in subsection~2.1.
In this equation, the magnetic field and gas pressures are considered to 
balance one another.  In fact, we can estimate the magnetic fields 
of Aql~X-1 and 4U~1608$-$52 by considering the cut-off luminosity, 
as indicated in  sub-subsection 2.5.1. 
The magnetic fields thus 
obtained are $ \sim 1.4 \times 10^8$ G for Aql~X-1 and $\sim 1.2 \times 
10^8$ G for 4U~1608$-$52, which correspond to a luminosity of 
$ \sim 1.6 \times 10^{36}$ erg s$^{^-1}$ for both sources as given in 
sub-subsection 2.5.1; note that we tentatively adopted 
$\eta^{-7/4}$ = 2 for the model dependence factor in equation (3), 
but further investigation on these magnetic fields will be published 
elsewhere in the future (Asai et al. in preparation).    
Here, we employed the pulse periods of 1.82 ms for  
Aql~X-1 (Zhang et al. 1998) and  1.62 ms for 4U~1608$-$52 
(Muno et al. 2001).

\begin{figure}
  \begin{center}
    \FigureFile(80mm,){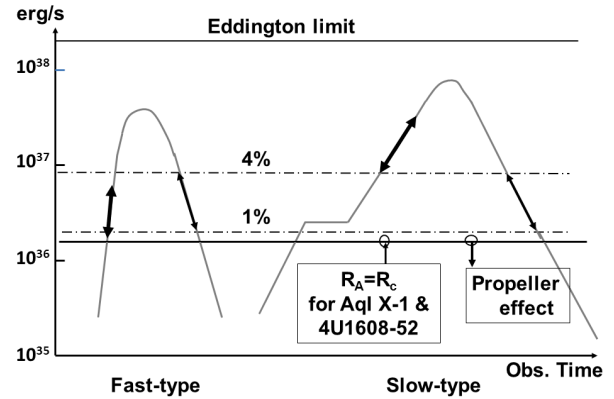}
  \end{center}
  \caption{Transition points along the light curves of two typical outbursts
of the fast rise and slow rise types (Asai et al. 2012).   The 
Eddington luminosity and fractional values of  1\% and 4\%  
of the luminosity are indicated by 
horizontal solid line and dashed-dotted lines, respectively.  Tentative 
luminosity values corresponding to $R_{\rm A} = R_{\rm c}$, at which 
the transitions from hard-high to hard-low states or vice versa 
will occur, are also indicated by horizontal lines for the two sources.
Since the transition from the hard-high state to soft state depends on 
the respective outbursts (Asai et al. 2012), the regions are indicated 
by arrows showing ranges. The transition from the soft state to the hard-high
state occurs somewhere between 1~ \% and 4~\%  the Eddington luminosity.  
These points are indicated by arrows as well.  
  
 }
\label{fig3}
\end{figure}

\subsubsection{Propeller effect at time of transition}
Figure 3 shows the relative luminosity levels of the three transitions
(hard-high state to soft state, the soft state 
to hard-high state,  and the hard-high state to hard-low state 
 and vice versa) for two outburst rise times. 
The nominal level of each transition is displayed along with the 
uncertainty derived from the result obtained by Asai et al. (2012), in which
we did not distinguish between the values for  Aql X-1 and 4U 1608$-$52,  
because they coincidentally exhibited similar behavior.  
The luminosity level at which $R_{\rm A} = R_{\rm c}$
is approximately given by  the point obtained from figure 2.

We consider that there are two types of mechanisms for 
plasma jets.  The first involves plasma ejection accompanying the
disc transition (the hard-high to soft transition, as mentioned in 
sub-subsection 2.3.3), 
whereas it is  uncertain for the time of the soft to hard-high transition. 
Another distinct mechanism occurs when plasma is ejected 
by the propeller effect when $R_{\rm A} > R_{\rm c}$.   The propeller effect 
in the outburst decay phase has been reported by
Campana et al. (1998a), Zhang et al (1998) and Chen, Zhang \& Ding (2006).  
However, none of these studies has distinguished between the  soft to 
hard-high and the hard-high to hard-low state transitions.  

It is thought  that in the decay phase the soft to hard-high disc transition 
occurs first, and then the hard-high to hard-low state transition follows.
If the gas accretion rate decreases rapidly, it is difficult to 
distinguish the two transitions.  Although 
an intensity gap (in particular, in a certain energy band) occurs sometimes  
in both transitions, the physical mechanisms differ.
The point at which $R_{\rm A} = R_{\rm c}$  corresponds to the threshold 
of the hard-high to hard-low transition; therefore, it is necessary to  
detect it first and then perform a careful analysis of which transition
occurred.  Campana et al (1998a) indicated a hard-high to hard-low state 
transition at a luminosity of $ \sim 4 \times $ $10^{36}$ erg s$^{^-1}$ 
for Aql X-1, whereas the result of Chen, Zhang \& Ding (2006) indicated
a luminosity of $(4.3 \sim 6.9) \times$ $10^{36}$ erg s$^{^-1}$ for 
4U~1608$-$52. Here, we assumed distances of 5.0 kpc and 4.1 kpc for 
Aql X-1 and 4U~1608$-$52, respectively (see Asai et al. 2012).
On the other hand Zhang, Yu, \& Zhang (1998) suggest $ \sim 10^8 $ G for Aql~X-1
(note that they assumed a distance of 2.5 kpc) and also $ 2 \times 10^9 $ G for Cen~X-4,
where they proposed a unified scheme for the spectral transition 
from the soft-high state to hard-low state without distinction of 
hard-high and hard-low states; note that their definition of states
differs from that in this paper.
These results differ from the values obtained in sub-subsection 2.5.2. 
Namely, both the present transition luminosities are  lower 
than the actual luminosities; consequently, the present magnetic fields
are  weaker than their results.

It is difficult to detect the intensity change due to the 
hard-high to hard-low state transition if it occurs 
at the same time as  the soft to hard-high transition.
However, it is not reasonable to expect that both transitions should occur simultaneously
for any source, because the luminosity level of the soft to hard-high transition 
is also common to BH-LMXBs (Maccarone 2003), and the hard-high to hard-low 
transition depends on the magnetic fields of individual neutron stars. 
In fact the difference between these intensity levels is seen in the MAXI/GSC 
light curves of Aql~X-1 and 4U~1608$-$52, although variability is large (see the MAXI  
home page : hppt://www.maxi.riken.jp/) 
as well as the intensity distribution in figure 2. the figure  implies 
a cut-off on the lower side of luminosity distribution (b).  
The luminosity below this cut-off is considered to correspond to 
that in the hard-low state.  Unfortunately, in figure 2 the lower part 
corresponding to the  hard-low state is  contaminated by considerable 
background.   Nevertheless, the X-ray luminosity in the hard-low state might be 
very greatly suppressed by the propeller effect; observational 
investigation of the suppressed quantity is a future issue.
Note here that X-ray emission in the quiescent period with a  
luminosity of  $ 10^{32 \sim 34}$ erg s$^{^-1}$ 
is often observed as discussed in sub-subsection 2.5.5.

On the other hand, the X-ray spectra before and after 
this transition indicate that the size of the blackbody region on the 
neutron star decreases
as the mass accretion rate decreases (Sakurai et al. 2011); i.e., the 
blackbody is smaller in the hard-low state than in the hard-high state. 
This indicates that in the hard-low state, the blackbody on the neutron 
star might be located only in the polar region.

\subsubsection{X-ray pulsations}  

In the hard-low state, gases from the companion star are accreted
around the magnetic poles of the neutron star, although considerable gas 
may instead be ejected by propeller effect.
This means that behavior of the neutron star is similar to that of an
X-ray pulsar, and a question arises as to why the pulse period of an NS-LMXB 
has not been easy to detect.  First, it is supposed that the rotation 
axis and the magnetic field axis of the neutron star are almost  
aligned, although this may not be  appropriate for an NS-LMXB that evolves
 into a milli-second pulsar.
Even if the blackbody regions near the poles vary in area,
detection of  a coherent pulsating signal will be difficult; however,
it may be possible to detect a QPO.
 Second, in addition to the weak intensity accreted on the polar 
region, the un-pulsed component from the inner region of the accretion disc also
degrades the signal-to-noise ratio.  Furthermore, the presence of a close 
companion star also disturbs simple pulsation.
Nevertheless, it is very important to investigate any pulsation
in the hard-low state of NS-LMXBs in future studies of the evolution of
neutron stars towards milli-second radio and/or gamma-ray pulsars from NS-LMXBs.  

Coherent milli-second pulsations in Aql~X-1 were observed for 150~s during 
bright X-ray emission, but not during an X-ray burst (Casella et al. 2008).
Since Aql~X-1 was in the soft state when  the  observations were made,
the presence of an equatorial belt needs to be accounted for in the pulsation study.
Thus, we prefer to use the ``burying'' scenario, in which a 
 certain variability in the magnetic field occurs (Casella et al. 2008).
The hot region of the neutron star is generally symmetric
about the spin axis.  However, if an asymmetric region were to be created,
it could last only for a short time; further,
the phase of the asymmetric region could not be constant during that 
time.  Although kHz-QPOs  typical of NS-LMXBs as well as Aql~X-1 
(Zhang et al. 1989) and 4U~1608$-$52 (Mendez et al. 1989) are often 
detected, the  observation of long-term coherent pulsation is a future 
issue.  

\subsubsection{X-ray spectrum in hard-low state}

The spectrum in the initial phase of the hard-low state, like that 
in the hard-high state, consists of 
thermal and non-thermal components associated with Comptonization.
The thermal spectrum is produced by the blackbody around the 
polar region and the multi-color blackbody of the accretion disc.
Because of the decrease in the accretion gas, the temperature of these 
blackbodies becomes lower; consequently, the Comptonized spectrum 
shifts to a non-thermal power law spectrum associated with 
the accreting gas near the neutron star.  This low luminosity NS-LMXBs 
are believed to exhibit the 
X-ray spectrum of a blackbody around the polar region and that 
defined by a non-thermal power law.    When the accretion becomes very 
weak, the blackbody may not be formed, and the gas  could hit the crust 
near the polar region instead, radiating the non-thermal power-law  
component.

X-ray observations performed during the quiescent periods of 4U~1608$-$52
and Cen~X-4 showed their X-ray luminosities to be on the order of
$10^{32-33}$ erg s$^{^-1}$ (Asai et al. 1996).
The spectra in the 0.5--10~keV range are fitted to a black body
model that incorporates a power law.
The radii of the blackbodies associated with 
4U~1608--52 and Cen~X-4 with the temperature of 
$0.2 \sim 0.3$~keV are estimated to be $\sim 1.5$~ km
and $\sim 1.8$~km, respectively. Aql~X-1 was also observed 
in its quiescent period using the ROSAT satellite, and 
its emission region was estimated to be of the order of 1 km$^2$ 
for a 0.3~keV blackbody spectrum (Verbunt et al. 1994).
However, these radii from data  obtained during the quiescent 
period have to be revised in light of recent investigations 
(Cackett et al. 2010, 2011; Fridriksson et al. 2011). 

Further X-ray observations during the quiescent period of Cen X-4 
have been made with Chandra, XMM-Newton and Suzaku 
(Cackett et al. 2010 and references therein).  The spectra thus obtained 
are fitted well by a model consisting 
of thermal emission ( from a 50$\sim $80 eV blackbody) 
with an additional hard-power-law tail, such that the thermal 
fraction of the spectra with variable intensity is 0.5$\sim$0.6.  The 
former component is presumably produced by  the neutron star's entire
surface, whereas the origin of the latter is unknown.  Aql~X-1 has been 
observed in the quiescent period  covering 2 years 
and exhibited X-ray variability at a luminosity of 
$(2 \sim 12) \times 10^{33}$ erg s$^{^-1}$ (Cackett et al. 2011). 
X-ray spectra are fitted to the thermal component  
$ kT = 110 \sim 120 $ eV and/or the power-law spectrum which are 
variable.  They suggest that the variability during quiscence is 
due to accretion at low rates that might reach the neutron star's 
surface.

A similar result was obtained in the quiescent period of a transient 
NS-LMXB, XTE J1701-462 by Fridriksson et al. (2011).  The 
thermal radiation on the neutron star's surface originating in crustal
 heating seems to deviate from the 
accreted radiation having a non-thermal spectrum. The effective 
temperature in the quiescent period of this source was 
$120 \sim 160$ eV with a luminosity of 
$(5 \sim 25) \times 10^{33}$ erg s$^{^-1}$.  The 
non-thermal component of a power law spectrum was also detected 
at 50\% or less of the total luminosity.  Note that the previous data
statistics of Asai et al (1996) were too poor to enable separation
 the blackbody from the power-law component.

Because the power-law spectrum is probably produced by 
weak accretion (Popham \& Sunyaev 2001), 
we might speculate that this component of the spectrum results 
from accretion of gas around the polar region during the quiescent 
period. In fact, as long as gas accretion  continues while 
$R_{\rm A}>R_{\rm c}$ and $R_{\rm A} < R_{\rm LC}$, this is regarded
as the hard-low state.  Namely, the power-law component could be 
produced by non-thermal gas around the polar region, whereas 
$ 60 \sim 160 $ eV blackbody radiation could be produced as the
neutron star's surface is heated by the  particles accreted 
around the polar region. Thus, this total
particle energy, which is not observable, could typically be 
larger than the luminosity of the power-law component. 
Then, a considerable amount of plasma (such as $ \sim 99\% $ of  the
mass flow rate) might be rejected by the 
neutron star, depending on the value of $R_{\rm A} $, as we also
discuss in  subsection 2.6.   Therefore, the above speculation is not
 appropriate  if the
behavior in the quiescent period occurs even  
when $R_{\rm A} > R_{\rm LC}$. 

\subsection{No-accretion X-ray state}

Although neither  Aql X-1 nor 41608$-$52 has been confirmed to be a radio 
milli-second pulsars, in this simplified picture, 
we adopt the recycled pulsar scenario according to which  NS-LMXBs 
will evolve to become  milli-second pulsars in  $\sim 10^9$ year 
(e.g., Stella et al. 1994; Wijnands \& van der Klis 1998;  Ho,
 Maccarone, \& Andersson 2011 and references therein).  
Although the evolution of milli-second pulsars has long been investigated 
 (e.g., Smarr \& 
Blandford 1976; Alpar et al. 1982), there has recently been a great 
deal of progress due to discoveries of a milli-second radio pulsar in a 
globular cluster (Lyne et al. 1987) and milli-second pulsations in
 the radio and/or X-ray bands of unidentified Fermi-LAT  
sources (Kerr et al. 2012; Kong et al. 2012; Bagchi, Lorimer, \& 
Chennamangalam 2011; Guillemot et al. 2012). Although much 
progress is being made in the comprehensive 
investigation on this problem, this topic is outside of the 
scope of the present paper.  

A recycled pulsar has a weak magnetic field and a period on
the order of milli-seconds.
Recycled pulsars are considered to be the link between NS-LMXBs
and radio or gamma-ray milli-second pulsars.
Thus, we define the no-accretion X-ray 
 state or the recycled pulsar state at a  time of no accretion 
 under   $R_{\rm A}>R_{\rm LC}$.  As some authors have 
recently suggested, this state may correspond to the quiescent state 
(Cackett et al. 2010, 2011; Fridriksson et al. 2011).
The X-ray intensity of transient NS-LMXBs often becomes very weak and 
is reduced to luminosities around $10^{32-34}$ erg $s^{-1}$.  
In  sub-subsection 2.5.5 we described this quiescent period as 
the hard-low state.  However, in Aql~X-1 and 4U~1608$-$52, 
these intensities are much lower than those corresponding to
$R_{\rm A} = R_{\rm LC}$ if we  neglect the 
propeller effect, where the luminosity is estimated 
to be  $ 10^{34 \sim 35}$ erg $s^{-1}$  (for $\eta = 0.5 \sim 1$) 
using  equation (4) in  subsection 2.1 
and the magnetic field obtained in sub-subsection 2.5.2.  

Therefore, the X-ray luminosity in the quiescent period  
of Cen~X-4 and XTE~J1701$-$462 as well as Aql~X-1 and 4U~1608$-$52 
in  sub-subsection 2.5.5
may correspond to the no-accretion X-ray state; in that case, non-thermal 
X-rays may be emitted from the boundary layer.  
Otherwise,  the  propeller effect is considered to account for as
much as  99 \%  or more of the mass flow rate 
from the companion star; cosequently, the quiescent period is still 
in the hard-low state.  However, it is necessary to investigate 
further whether the observed quiescent state (we called this 
state ``period''  in this paper because it is a physically unestablished 
state) belongs to the hard-low state or the no-accretion X-ray state.

In this situation there may be no clear transition in the X-ray band 
at $R_{\rm A} = R_{\rm LC}$ although the physical mechanism  
changes at this point.  Nevertheless, it is important to detect this point
in order to understand the recycled pulsar scenario.
If there is no further increase in the accretion rate, a recycled pulsar will
sometimes be born accompanying  the radiation from the light cylinder barrier.
Some radiation may be produced
around the light cylinder or  
polar region because of slight leakage of accretion gas at 
$R_{\rm A}>R_{\rm LC}$.  These are problems to be solved in the future.

\section{Discussion and summary}

\subsection{Difference in spectra depending on viewing angle}

We have proposed a simplified picture of NS-LMXBs based on the recent 
results of two soft transients, Aql~X-1 and 4U~1608$-$52
(e.g.; Asai et al. 2012 and references therein).  For these two
NS-LMXBs, the X-ray spectra in their soft state exhibit a 
thermal spectrum and a weak Comptonized component, whereas the 
spectra for the hard-high state exhibit mainly Comptonized spectra 
extending to $\sim ~100$ keV in addition to seed blackbody components.
In the same state, the spectra resemble  each other, but 
some difference is expected when  detailed spectral observations 
are made.  No dip-like detection has not been reported in 4U~1608$-$52, 
whereas dipping activity was detected 
in the outburst of Aql~X-1, although infrequently (Galloway 2012).
Therefore, some fraction of NS-LMXBs exhibit such dipping behavior.
Some  dipping NS-LMXBs do not exhibit the hard component extending to 
$\sim100$~ keV even in the hard state (Sugizaki et al. 2012); for example,
the energy of Comptonization electrons determined through  a fitting parameter
is on the order of several keV for MAXI J0552-332 (a dipping source: Strohmayer 
\& Smith 2011), compared to several tens of keV in the hard states
of Aql~X-1 and 4U~1608$-$52.  We discuss this behavior using the  
conceptual cartoon in figure 4, 
although a detailed investigation of spectral models is not the topic of this study.

We consider the idea that this difference can originate because of the
viewing angle of the accretion disc surface.
We assume that there are generally two types of electron gas energies
responsible for Comptonization: high energy electron gases (on
the order of several tens of~keV) and low energy electron gases
(on the order of several keV).
Although an actual plasma with high energy electrons may be continuously
distributed in energy, we consider a simplified picture with 
representative electrons having only two types of energy.
The high energy plasma with electrons having energies of several 
tens of keV  is concentrated near the neutron star's surface,
whereas the low energy plasma with electrons having energies of several  
keV is located outside of the high energy plasma and  limited to the 
area near the disc.
The  high energy plasma is larger than  the low energy plasma.

In the hard-high state, we usually observe  hard X-rays Comptonized by 
the high energy electrons for a small viewing angle;  4U~1608$-$52 
corresponds to this case, whereas the occurrence of occasional dipping
indicates that Aql~X-1 has a slightly large viewing angle.  On the other hand, if
the viewing angle is large (i.e., a large inclination or a dip source),
we observe most X-rays through the low energy plasma,
where we have speculated that most seed photons are provided from
the blackbody on the neutron star and the inner region of the multicolor disc.
Therefore, even if the X-rays from a neutron star
 consist of hard X-rays at energies of several tens of keV, the energy
of the observed X-rays will be shifted to a lower energy due to Comptonizxation
by the low energy plasma with several keV electrons. Therefore,  
the spectrum of Aql~X-1 is expected to be slightly softer than that of 
4U~1608$-$52, considering the dipping behavior  of Aql~X-1.  However, 
further investigation of this question is a future issue.

Here, we note the relationship between the above hot plasma and the accretion
disc corona (ADC: e.g., Church \& Baluci\'{n}ska-Church 2004).  In the present
picture, we consider that the hot plasma is created near the neutron
star and spreads only to the nearby region of the disc, whereas the ADC 
has lower temperature than the hot plasma and spreads to the overall disc, 
thus causing a dip for some NS-LMXBs with a very large viewing angle.
The location of these plasmas is illustrated in figure~4.

Dipping sources are usually not as simple (Hirano et al. 1995; 
Church \& Baluci\'{n}ska-Church 
2004 and references therein; Church, Jackson \& Baluci\'{n}ska-Church 2008). 
Thus, the geometrical situation of the electron gases associated with  
Comptonization complicates the X-ray spectra  because of the time variability  
of the hot plasma and the ADC depending on the source as well as the mass  
accretion rate.  In fact, the variable nature of these plasmas would cause 
NS-LMXBs with large inclination to have complex spectra.
If we consider the distinctiveness of the hot plasma distribution, the present 
simplified picture near the neutron star might be
saved in principle and thus, be applicable at any viewing angle.

\begin{figure}
  \begin{center}
    \FigureFile(80mm,){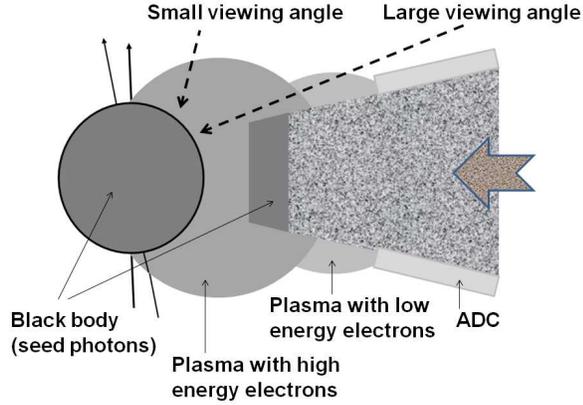}
  \end{center}
  \caption{Conceptual cartoon indicating the spectral difference due to different
viewing angles [in case of hard-high state (2) in figure~1].
ADC~=~accretion disc corona (see the text in subsection 3.1).}
\label{fig4}
\end{figure}

\subsection{Correlation between present representation and color--color 
and hardness--intensity diagrams}

We have tried to simplify the physical picture of NS-LMXBs with respect to the
intensity evolution from the brightest state to the weakest state.
Each state exhibits a spectrum with specific physical characteristics.
Since the X-ray intensity varies widely in both the soft and hard states,
it is generally difficult to classify these states based on the hardness/softness
ratios. However, we attempt to correlate the present simplified picture 
with the color--color and hardness--intensity diagrams, although these two 
paradigms are independent of each other. 
In figure 5, we roughly indicate the soft, hard-high, 
hart-low states and the hard-to-soft or soft-to-hard transition 
regions in the color--color and hardness--intensity diagrams for the 
two sources: the original diagrams are reproduced from figure 2 of the paper 
by Lin, Remillard, \& Homan (2007).  Note that the 
hard-low state and the transition from the hard-high state to the hard-low state 
are not obvious in this figure.   
 
\begin{figure}
  \begin{center}
    \FigureFile(80mm,){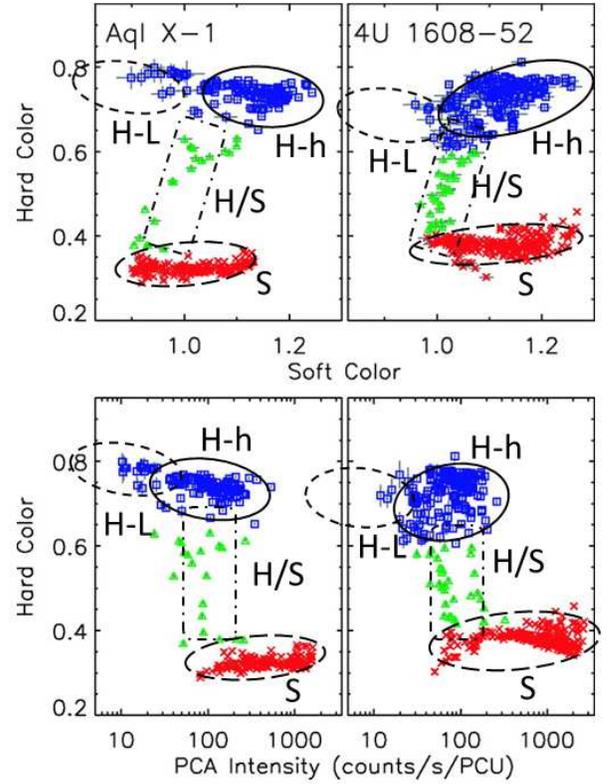}
  \end{center}
  \caption{
Conceptual states and state transitions on typical   color--color 
diagrams (upper panel) and  hardness--intensity diagrams (lower panel) for Aql~X-1 
and 4U~1608--52 reproduced from figure 2 by  Lin, Remillard, \& Homan (2007).  
S: soft state, H-h: hard-high  state, H-L: hard-low sate 
H/S: soft to hard-high transition and vice versa. Hard-high to 
hard-low transition will occur somewhere between the hard-high and hard-low states. 
}
\label{fig5}
\end{figure}

NS-LMXBs have been classified as Z-sources and Atoll-sources
using X-ray color--color diagrams (Hasinger \& van der Klis 1989)
and hardness--intensity diagrams (Schulz, Hasinger \& Truemper  1989).
They have also been studied in order to examine the characteristics
of QPO behavior.  Z-sources remain primarily in the soft state,  
whereas the Atoll-sources, Aql~X-1 and 4U~1608$-$52, have been observed  
in both the soft and the hard states.
The three transitions mentioned above can be observed in
the Atoll-sources, but the hard-high and hard-low states are not clearly  
distinguished in the color--color and hardness-intensity diagrams because of 
the data selection effect such as that in the hard-low data in Lin, Remillard, 
 \& Homan (2007).  Note that MAXI/GSC has provided lower luminosity data for Aql~X-1 
and 4U~1608$-$52,  as shown in figure 2.  In contrast, the  transitions from
the soft state to the hard-high state and vice versa are roughly apparent 
in the diagrams.
In particular the hardness and softness ratios cause interference
between the hard-high and hard-low states because these states share
the energy of 2--20~keV, along with other narrower bands.
Therefore, in order to understand overall physical characteristics
of NS-LMXBs we require further information for extended hard X-rays over
at least $\sim 100$~keV as well as long-term monitoring. 

\subsection{Summary of proposed simplified picture}

Before summarizing our simplified picture of the NS-LMXBs Aql~X-1
and 4U~1608$-$52, we  emphasize that it  differs from 
 previous unified schemes (e.g., Campana et al. 1998b; Zhang, Yu \& 
Zhang 1998).  Our scheme is based on the mass flow rate and  magnetic field,
so it depends on the balance of $ R_{\rm c}, R_{\rm A}$ and $R_{\rm LC} $,  
as indicated in figure 1.  Using this scheme we introduced 
the hard-high and hard-low states which have not been distinguished previously.  
This introduction at  $R_{\rm A}=  R_{\rm c}$  is natural for  neutron stars 
 with a considerably strong magnetic field. We identified this point probably in the MAXI
observation data shown in figure 2.  Therefore, at least three states and
three state transitions and another state transition point of 
$R_{\rm A}=  R_{\rm LC}$ might be a reasonable scheme for 
NS-LMXBs in general, but a fourth state, the no-accretion X-ray state, is necessary
for comprehensive investigations in the future. 

We summarize as follows.
The NS-LMXB spectrum in  the soft state exhibits a major blackbody 
and multi-color disc emission component, and a minor 
Comptonized component.  The spectrum in the hard-high state is considered 
to be due mainly to  Comptonization, in which the blackbody region covering 
the neutron star's entire surface and the inner blackbody region of the multi-color 
disc may produce seed photons.
The difference between  the Comptonized emissions in the soft and hard-high states
is due to a difference in the size of the hot plasma accreting from the 
corresponding accretion discs.  The  hot plasma is smaller in the soft state than
in the hard-high state.  In Aql~X-1 and 4U~1608$-$5, 
there is a large degree of variability 
in the hard-high state, which tends more toward hard X-rays, than  
in the soft state, if the luminosity in the soft state is
not above or near the Eddington limit. 
The variability in the hard-high state is reduced toward the hard-low state 
according to the decrease in the mass flow rate from the companion star.

In the hard-low state, the mass flow rate decreases, and the amount 
of seed photons also decreases accordingly.
Although the gases will accrete widely over the entire neutron star,
considerable gas might be ejected by propeller effect, and  
some gases eventually accrete around both magnetic polar regions of the neutron 
star.  In this state, the  Comptonized radiation will decrease owing to 
the decreasing gas density and decreasing blackbody area, which leads to the
production of seed photons.  Thus, a spectrum with weak luminosity eventually 
transitions to that of a small blackbody along  with a power law spectrum.
Note that the ratio of the accreted gases to the ejected gases is not
certain at present.  This problem should be investigated by the studying
the quiescent X-ray emission in the no-accretion
X-ray state as well as the hard-low state. 

Concerning  the no-accretion X-ray state, the accretion X-ray emission will 
probably stop when
the Alfv\'{e}n radius is comparable to or larger than that of the light cylinder, 
but it is necessary to investigate this state further; e.g., in terms of the 
recycled pulsar scenario and  radio and gamma-ray observations of milli-second 
pulsars.

Our picture also suggests four main state transitions:
from the hard-high state to the soft state,
from the soft state to the hard-high state,
from the hard-high state to the hard-low state, or vice 
versa, and from the hard-low state to the  no-accretion X-ray state or
 recycled pulsar state,  or vice versa, although the last transition 
is not yet confirmed observationally.
These transition points cannot be attributed strictly
to the gas accretion rates from the companion star because of the
hysteresis effect; in particular, it is quite
large during the transition from the hard-high state to the soft state.  

The effectiveness of our simplified picture can be confirmed by observing 
 recurrent NS-LMXB that ranges from the weakest to the strongest
intensities, if the wide band spectra of such  NS-LMXBs can be
continuously observed using instruments with good sensitivity.
On the other hand, because a steady NS-LMXB remains in a certain state without
a state transition for a long time, we must select only  the state for which
 a careful analysis can be conducted  using wide-band spectral observations.

\bigskip
We would like to thank Profs. H.Inoue,  K.Makishima, R.Matsumoto and S.Mineshige 
for their stimulated comments and discussions on this work. We also acknowledge 
Dr.T.Mihara and an anonymous referee deeply for their useful comments and instructive advice.

\end{document}